\documentclass[fleqn,11pt]{article}

\usepackage{amssymb,amsmath}

\usepackage{amsfonts,amssymb,cite}
\usepackage{graphicx}

\def\noi{\noindent}

\newcommand{\Title}[1]{\noi {{\Large\bf #1}}\\[1ex]}

\def\Aunames#1{\noi{\bf #1}}

\def\Addresses#1{\medskip\noi \protect
	\begin{description}\itemsep -3pt {\it #1} \end{description}}
\def\adr#1#2{\item[${}^{#1}$]{\it #2}}

\newcommand{\Abstract}[1]{\vskip 2mm \begin{center}
        \parbox{16.4cm}{\small\noi #1} \end{center}\medskip}

\def\email#1#2{\footnotetext[#1]{e-mail: #2}\addtocounter{footnote}{1}}

\def\nqq{\hspace*{-2em}}

\usepackage{color}

\def\Jl#1#2{#1 {\bf #2},\ }

\def\ApJ#1 {\Jl{Astroph. J.}{#1}}
\def\CQG#1 {\Jl{Class. Quantum Grav.}{#1}}
\def\DAN#1 {\Jl{Dokl. AN SSSR}{#1}}
\def\GC#1 {\Jl{Grav. Cosmol.}{#1}}
\def\GRG#1 {\Jl{Gen. Rel. Grav.}{#1}}
\def\IJMPD#1 {\Jl{Int. J. Mod. Phys. D}{#1}}
\def\JETF#1 {\Jl{Zh. Eksp. Teor. Fiz.}{#1}}
\def\JETP#1 {\Jl{Sov. Phys. JETP}{#1}}
\def\JHEP#1 {\Jl{JHEP}{#1}}
\def\JMP#1 {\Jl{J. Math. Phys.}{#1}}
\def\NPB#1 {\Jl{Nucl. Phys. B}{#1}}
\def\NP#1 {\Jl{Nucl. Phys.}{#1}}
\def\PLA#1 {\Jl{Phys. Lett. A}{#1}}
\def\PLB#1 {\Jl{Phys. Lett. B}{#1}}
\def\PRD#1 {\Jl{Phys. Rev. D}{#1}}
\def\PRL#1 {\Jl{Phys. Rev. Lett.}{#1}}

\def\lal{&&\nqq {}}

\def\beq{\begin{equation}}
\def\eeq{\end{equation}}
\def\bear{\begin{eqnarray}}
\def\bearr{\begin{eqnarray} \lal}
\def\ear{\end{eqnarray}}
\def\earn{\nonumber \end{eqnarray}}

 \catcode`\@=11 \@addtoreset{equation}{section}\catcode`\@=12

\begin{document}
\twocolumn[

%\jnumber{4}{2025}

\Title{Photon Sphere for a Dilatonic Dyonic Black Hole in a Model
       with an  Abelian Gauge Field and a Scalar Field   }

\Aunames{V. D. Ivashchuk$^{a,b,1}$,  U. S. Kayumov$^{c}$, A. N. Malybayev$^{c}$, and  G. S. Nurbakova$^{c}$, } 

\Addresses{
\adr a {\small Peoples' Friendship University of Russia (RUDN University), 
             ul. Miklukho-Maklaya 6, Moscow 117198, Russia} \\         
\adr b {\small Center for Gravitation and Fundamental Metrology, SRCAM Rostest,
            ul. Ozyornaya  46, Moscow 119361, Russia} \\
\adr c {\small 
    Al-Farabi Kazakh National University,  Al-Farabi avenue, 71, Almaty 050040, Kazakhstan } 
	}

%\Dates{\ xx, 2025}{\ yy, 2025}{\ zz,  2025}

\Abstract
{
Dilatonic dyon black hole  solution with gravitational radius $2 \mu$ and two  charges $Q_1$ and $Q_2$ (electric
and magnetic ones) in the gravitational $4d$ model with one scalar field  
and one 2-form is considered. Dilatonic coupling constant $\lambda$ obeys $\lambda^2 = \frac{1}{2}$.
The circular orbits for null geodesics are explored. The 3rd order polynomial master equation for radius
 $R_0$ of  photon sphere is  studied. It has only one solution which obeys $R_0 > 2 \mu$.
 The circular null geodesics are shown to be unstable. The  black hole shadow is studied 
 and relations for shadow angle and critical impact parameter are obtained.
     
}

] %%%%%%%%%%%%%%%%%%%%%%%%%%%%%%%%%%%%%%%%%%%%%%%%%%%%

\email 1 {ivashchuk@mail.ru}

{ % au-def
\def\R{{\mathbb R}}

% ====================================
%%%%%%%%%%%%%%%%%%%%%%%%%%%%%%%%%%%%%%%%%%%%%%%%%%%%%%%%%%%%%%%%
\section{Introduction}
%%%%%%%%%%%%%%%%%%%%%%%%%%%%%%%%%%%%%%%%%%%%%%%%%%%%%%%%%%%%%%%%

Interest in black holes, originally inspired by evidence for a supermassive black hole in our galactic center \cite{Ghez}, has been strengthened by the recent discovery of gravitational waves emitted from a black hole merger \cite{Abbott}. Many current papers consequently investigate black holes in modified theories of gravity (reviewed in \cite{Vagn} and references therein). In this work, we consider one such theory - 
a dilatonic scalar-tensor model with one scalar field and one Abelian gauge field.

This paper builds upon previous research \cite{ABDI,ABI,BBIM,MBI} 
focused on dilatonic dyonic and dyon-like black hole solutions. 
These findings can be regarded as a small component 
of the vast body of work addressing spherically symmetric solutions, 
such as black holes and black branes, 
as discussed in \cite{BronShikin} - \cite{GKO} and the references therein. 
Such solutions emerge in gravitational models that incorporate scalar fields and 
 antisymmetric  forms.

Here we consider dilatonic dyon black hole solution 
\cite{ChHsuL} with  electric and magnetic
charges $Q_1$ and $Q_2$, respectively, in the $4d$ model with metric $g$, 
one scalar field $\varphi$ and one 2-form  $F$, 
corresponding to   dilatonic coupling constant  
\beq \label{i4.1}
  \lambda  = \pm 1/\sqrt{2}.
 \eeq

 For a dyon configuration on an oriented manifold 
${\cal M } = (2\mu, + \infty)  \times S^2 \times  \R $ ($\mu > 0$), considered
below, the 2-form field must satisfy the following relation
\beq 
 F = Q_1 f * \tau + Q_2 \tau, \label{0.1}
 \eeq
where  $\tau = {\rm vol}[S^2]$  is volume form on $2d$ sphere, $* = *[g]$ is the Hodge operator for
 $({\cal M }, g)$ and $f=f(x)$ is a certain function on ${\cal M }$. 
 
 In Ref. \cite{ABI}, the term ``dyon-like'' configuration was introduced to differentiate 
 the non-composite ansatz from the composite dyonic one given in equation (\ref{0.1}).
 In this case we have two 2-forms $F^{(s)}$, $s =1, 2$, with the ansatz: 
    $F^{(1)} = Q_1 f * \tau$,  $F^{(2)} = Q_2 \tau$, imposed instead of (\ref{0.1}).

 In the context of two Abelian 2-form fields, specific dyon-like solutions associated 
with various Lie algebras and dilatonic couplings $\lambda_1, \lambda_2$  
were investigated in Refs. \cite{GM,KLOPP,ABI,Dav,GalZad,AIMT}. 
The case involving the  $A_1 + A_1$  Lie algebra, which corresponds to the condition 
$\lambda_1 \lambda_2 = \frac{1}{2}$, was examined in Refs. \cite{GM,KLOPP,ABI,AIMT,IMNT}. 
For $\lambda_1 = \lambda_2 = \lambda$ we are led to relation (\ref{i4.1}) which appears in 
certain $D =4$ supergravity  and stringy induced models \cite{GM,KLOPP}.
Dyonic (composite) solutions have been the subject of numerous studies, 
as referenced in  \cite{Lee,ChHsuL,GKLTT,PTW,Br0,FIMS,GKO,ABDI} and related works.

In fact the dyonic solution from Ref. \cite{ChHsuL} (see also \cite{Br0})  
is a non-composite version of the solutions from Refs. \cite{GM,KLOPP}.

In this paper, we investigate the circular null geodesics at a constant radius 
$R = R_0$, 
which coincides with the radius of the photon sphere. 
The parameters associated with photon spheres are crucial for various applications 
of black hole solutions, including the spectra of quasinormal modes (QNM) 
in the eikonal approximation, circular orbits for massive particles, 
black hole shadows, and more, as discussed in 
Refs. \cite{Vagn,KZh,CGP,KSt,BTIMNU} and the references therein. 
  
The paper has the following structure. 
In Section 2 we describe the model and  the dyon black hole solution. 
In Section 3 we present  certain physical parameters of dyon black hole (BH).  
In Section 4, we derive the master equation for the radius $R_0$ of a circular photon orbit 
within the background metric of the BH dyon solution and prove a proposition concerning the existence 
and uniqueness of the solution to the master equation that satisfies $R_0 > R_g$, 
where $R_g = 2 \mu$ is the horizon radius.
Here we also prove the instability of circular photon orbits.   
In Section 5 we consider the BH shadow. 

%%%%%%%%%%%%%%%%%%%%%%%%%%%%%%%%%%%%%%%%%%%%%%%%%%%%%%%%%%%%%%%%
\section{Black hole dyon solution}
%%%%%%%%%%%%%%%%%%%%%%%%%%%%%%%%%%%%%%%%%%%%%%%%%%%%%%%%%%%%%%%%

Let us consider a model governed by the action

\begin{eqnarray}
 S= \frac{1}{16 \pi G}  \int d^4 x \sqrt{|g|}\biggl\{ R[g] -
   g^{\mu \nu} \partial_{\mu} \varphi  \partial_{\nu} \varphi
   \label{i.1}  \\
 - \frac{1}{2} e^{2 \lambda \varphi} F_{\mu \nu} F^{\mu \nu }
 \biggr\},   \nonumber 
\end{eqnarray}
where $g= g_{\mu \nu}(x)dx^{\mu} \otimes dx^{\nu}$ is  metric,
 $\varphi $ is the  scalar field, 
 $F = dA  =  \frac{1}{2} F_{\mu \nu} dx^{\mu} \wedge dx^{\nu}$
is the $2$-form with $A = A_{\mu} dx^{\mu}$; 
$G$ is the gravitational constant;
 $\lambda$ is  coupling constant  
 obeying (\ref{i4.1})  and  $|g| =   |\det (g_{\mu \nu})|$. 
 
 We consider  dyon black hole
solution \cite{ChHsuL,ABDI} to the field equations corresponding to the action
(\ref{i.1})  which is defined on the manifold
\beq \label{i.2}
 {\cal M }  =    (2\mu, + \infty)  \times S^2 \times  \R,
\eeq
and has the following form
\bear  \nonumber
 ds^2 = g_{\mu \nu} dx^{\mu} dx^{\nu} \qquad \qquad
 \\ \nonumber
 = H_1 H_2
 \biggl\{ -  H_1^{-2 } H_2^{-2 } 
 \left( 1 - \frac{2\mu}{R} \right)
 dt^2 \\ 
  \qquad +  \frac{dR^2}{1 - \frac{2\mu}{R}} + R^2  d \Omega^2_{2}
  \biggr\},  \label{i.3}
 \\  \label{i.3a}
 \exp(\varphi)=
 H_1^{\lambda } H_2^{- \lambda},
 \\  \label{i.3bem}
 F=  \frac{Q_1}{R^2}   H_{1}^{-2}  dt \wedge dR
     +  Q_2 \tau.
\ear

Here  $Q_1$ is electric charge and $Q_2$ is magnetic charge, 
$\mu > 0$,  $d \Omega^2_{2} = d \theta^2 + \sin^2 \theta d \phi^2$
is the  metric on the unit sphere $S^2$
 ($0< \theta < \pi$, $0< \phi < 2 \pi$),
 $\tau = \sin \theta d \theta \wedge d \phi$
is the  volume form on $S^2$. 
For the speed of light we put $c=1$.

The functions $H_s$ are given by
\beq \label{i4.2}
 H_s = H_s(R)  = 1 + \frac{P_s}{R},
\eeq
where
\beq \label{i4.3}
 P_s (P_s + 2 \mu) = Q_s^2,
\eeq
$s = 1,2$. By utilizing the  positive roots
from Eq. (\ref{i4.3})
\beq \label{i4.3p}
    P_s  = P_{s,+} =  - \mu + \sqrt{\mu^2 +  Q^2_s} > 0,
\eeq
we arrive at a well-defined solution  for $R > 2\mu$.

The moduli functions $H_s$ satisfy 
the following boundary conditions:
\beq \label{i3.1a}
  H_s  \to H_{s0} = 1 + \frac{P_s}{2 \mu}  > 0
\eeq
for $R \to 2\mu $, and
\beq \label{i3.1b}
  H_s    \to 1
\eeq
for $R \to +\infty$, $s = 1,2$.

In brane terminology, this is a composite solution that describes a configuration 
of two intersecting non-extremal black $0$-branes - an electric one 
and a magnetic one - with the intersection rule corresponding 
to the Lie algebra $A_1 + A_1$ ($A_1 = sl(2)$) \cite{IMtop,Isym}.

According to the first boundary condition (\ref{i3.1a}), we obtain a (regular) horizon at 
$R = R_g = 2 \mu$ for the metric (\ref{i.3}). The second condition 
(\ref{i3.1b}) guarantees asymptotic flatness of the metric as 
$R \to +\infty$.

 A global extension of the metric (\ref{i.3}), initially defined for 
 $R > 2 \mu$, reveals the presence of two horizons at $R = 2 \mu$ 
 and $R = 0$,  and a singularity at $R = - \min (P_1,P_2)$.
  We note that for $P_1 = P_2$ the  metric (\ref{i.3}) 
 is coinciding with the  Reissner-Nordstr\"om metric.

\section{Physical parameters}

Here, we provide specific physical parameters associated with the black hole solution.

\subsection{Gravitational mass and scalar charge}

From equation (\ref{i.3}), we find that the ADM gravitational mass is given by
 \beq \label{i5.1}
 GM =   \mu +  \frac{1}{2} (P_1 + P_2),
\eeq
 where $G$ represents the gravitational constant.

The scalar charge is simply derived from equation (\ref{i.3a})
\beq \label{i5.1s}
 Q_{\varphi} =  \lambda (P_1 -   P_2).
\eeq

Using the relations (\ref{i4.3}), (\ref{i5.1}), and (\ref{i5.1s}), 
we obtain the following identity
 \beq \label{i5.1id}
     2 (GM)^2   +     Q_{\varphi}^2   = Q_1^2 + Q_2^2 + 2 \mu^2.
 \eeq
In the extremal case $\mu = +0$ this relation 
 was found earlier  in \cite{PTW}.

\subsection{The Hawking temperature and  entropy}

  The Hawking temperature, calculated from the solution, 
  is  \cite{ABDI}
  \beq \label{i5.2}
 T_H=   \frac{1}{8 \pi \mu}  H_{10}^{- 1} H_{20}^{- 1}.
 \eeq
Here $H_{s0}$ are defined in (\ref{i3.1a}) and 
 relations $c = \hbar = k_B = 1$ are adopted.

For the Bekenstein-Hawking (area) entropy $S = A/(4G)$ ($A$ is the horizon area),
corresponding to the horizon at $R = 2\mu$, we obtain
\beq \label{i5.2s}
S_{BH} =   \frac{4 \pi \mu^2}{G}  H_{10} H_{20}.
\eeq
Relations (\ref{i5.2}) and (\ref{i5.2s}) imply 
%\beq \label{i5.2st}
 $T_H  S_{BH} =   \frac{\mu}{2G}$.
%\eeq

\section{Circular geodesic solutions}

In this section, we investigate the geodesic equations associated with the metric (\ref{i.3}). 
Our focus is on circular solutions for both null and time-like geodesics.

\subsection{Geodesic equations}

The geodesic equations can be obtained from the Lagrangian
\beq \label{4.Lag}
    \mathcal{L} = \frac{1}{2}  g_{\alpha \beta}(x) \dot{x}^{\alpha}\dot{x}^{\beta}.
\eeq
They are equivalent to the Euler-Lagrange equations
\beq \label{4.ELeqs}
    \frac{d}{d\tau}\left(\frac{\partial \mathcal{L}}{\partial \dot{x}^{\alpha}}\right)
     - \frac{\partial \mathcal{L}}{\partial x^{\alpha}} = 0.
\eeq
Here $\dot{x}^{\alpha}=dx^{\alpha}/d\tau=u^{\alpha}$ is the 4-velocity vector 
($x^{\alpha} = x^{\alpha}(\tau)$, $\alpha=0, 1, 2, 3$), 
where $\tau$ represents the proper time for a massive point-like particle 
moving along a timelike geodesic, and serves as the affine parameter 
in the case of a null geodesic, respectively.

We normalize 4-velocity vector  as following 
\beq \label{4.normal}
     g_{\alpha \beta}(x)  u^{\alpha} u^{\beta} = -k = 2 {\cal E},
\eeq
where $k=-1,0,1$ for spacelike, null and timelike geodesics, respectively. 
The quantity ${\cal E}$ 
is the energy integral of motion for the Lagrange equations, corresponding 
to Lagrangian (\ref{4.Lag}). The spacelike geodesics are irrelevant for our consideration.

For the redshift function and central function, we adopt the notations 
$A(R)$ and $C(R)$, respectively,
\bear
A =A(R) = H_1^{- 1} H_2^{- 1} \left(1-\frac{2\mu}{R}\right),
    \label{4.A} \\   
C =C(R) = H_1 H_2 R^2.      \label{4.C}
\ear

Without loss of generality, we consider null or timelike geodesics in the equatorial plane 
($\theta = \pi/2$). This effectively reduces the system to a 3D problem, for which the 
Lagrangian derived from the metric in Eq. (\ref{i.3}) reads:

\beq \label{4.Lagrangian}
     \mathcal{L}_{*}
    = - A(R) \dot{t}^2 + (A(R))^{-1} \dot{R}^2 + C(R) \dot{\phi}^2.
\eeq

This Lagrangian governs the geodesic equations corresponding to the variables  $t$, $R$ and $\phi$, 
while the geodesic equation for the $\theta$ variable is satisfied identically 
for our adopted choice of $\theta = \pi/2$.

The cyclic coordinates $t$ and $\phi$ yield the following integrals of motion:
\beq \label{4.ELhat}
  \hat{E} = A(R) \dot{t},\qquad  \hat{L} = C(R) \dot{\phi}.
\eeq
For $k=1$, they are related to the total energy $E = \hat{E} m$ 
and angular momentum $L = \hat{L} m$ of a test (neutral point-like) particle with mass $m > 0$.

Equation (\ref{4.normal}) takes the following form for the line element given in  (\ref{i.3})
\beq \label{eq:k}
    -  A(R) \dot{t}^2 + 
       (A(R))^{-1} \dot{R}^2 + C(R) \dot{\phi}^2 = -k. 
\eeq
Using Eqs. (\ref{4.ELhat}), we derive the following differential equation  
\beq \label{4.EqWithEL}
    -\frac{ \hat{E}^2}{A} + 
    \frac{ \dot{R}^2}{A} + \frac{\hat{L}^2}{ C} = -k,
\eeq
which can be reformulated in terms of the effective potential
\beq \label{4.EqForR}
    \dot{R}^2 + U(R) = \hat{E}^2,
\eeq
explicitly given by
\bear \label{4.eff_pot}
    U = U(R) = A(R) \left( k +  \frac{\hat{L}^2}{ C(R) }   \right) \qquad  \\ \nonumber
    = H_1^{-2} H_2^{-2} \left(1-\frac{2\mu}{R}\right)
    \left(H_1 H_2 k + \frac{\hat{L}^2}{ R^2}\right).
\ear

  It is easy to verify that the Lagrange equation for the radial coordinate $R$ is expressed as follows:
 \beq \label{4.EqForRtrue}
   2 \ddot{R} +  \frac{\partial U}{\partial R} = 0.
\eeq
 For $\dot{R} = 0$, the radial equation reduces to
 %\beq \label{4.EqForR0}
    $\frac{\partial U}{\partial R }= 0$.
%\eeq
%It does not follow from Eq. (\ref{4.EqForR}) and should be considered separately.

\subsection{Circular photon orbits}

 We will now address circular solutions for the null geodesic equations that satisfy:
\beq \label{4.R0}
  R (\tau) = R_0 = {\rm const}, \qquad R_0 > 2 \mu.
 \eeq 
 The parameter $R_0$ is referred to as the radius of the photon sphere which 
 is covered by  circular photon orbits.
 
 In this case $\hat{L} \neq 0$ and the radial equation (\ref{4.EqForRtrue}) for 
 $R = R_0$  with $k= 0$ reads 
 as $(A/C)' =0$, where prime means a differentiation over $R$, or, equivalently,
    \begin{eqnarray}           
      R^3  - 3 \mu  R^2  \nonumber \\
         + [ \mu ( -  P_1 -  P_2)  - P_1 P_2 ] R   +  \mu  P_1 P_2  = 0. 
           \label{4.masteqR}
   \end{eqnarray}

 {\bf Proposition 1.} {\it For all $\mu > 0$, $P_1 > 0$ and  $P_2 > 0$ 
the third order polynomial (master) equation  (\ref{4.masteqR})
has one and only one real solution $R_0$ which satisfies the inequality $R_0 > 2 \mu$}.

In other words the Proposition 1 states the existence and uniqueness of photon 
sphere (outside the event horizon) for the metric (\ref{i.3}) 
for any proper set of parameters.

For the sake of convenience we rewrite the master 
equation  (\ref{4.masteqR}) in terms of dimensionless parameters
\begin{equation} 
  \label{4.xp}
  x = R/(2 \mu), \qquad p_i = P_i/(2 \mu),
\end{equation}
$i = 1,2$. 

We get 
  \begin{eqnarray}   \nonumber     
   P(x) = x^3 - \frac{3}{2} x^2 
   \qquad \qquad \qquad \qquad  \\ 
   + \left(  -  \frac{1}{2} p_1 - \frac{1}{2} p_2 -  p_1 p_2 \right) x   +  \frac{1}{2}  p_1 p_2  = 0.
   \qquad   \label{4.masteqx}
 \end{eqnarray}  

 Graphical representation of the function $P(x)$ for $p_1 = p_2 = 1$ is given at Figure 1.

 \begin{figure}[h]
 \center{\includegraphics[scale=0.4]{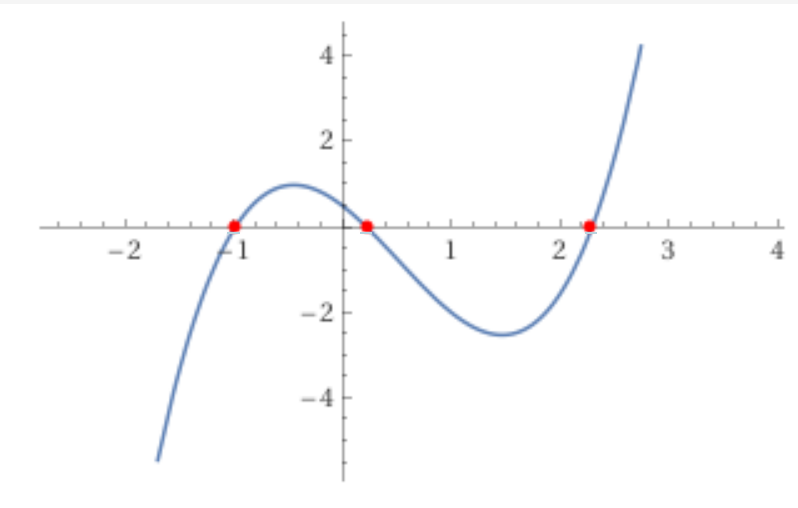}}
 \caption{The function $P(x)$ from (\ref{4.masteqx}) for $p_1 = p_2 = 1$.  } 
 \label{Fig_1}
 \end{figure}
 
The Proposition 1 is valid due to the following proposition.
 
{\bf Proposition 2.} {\it For all  $p_1 > 0$,  $p_2 > 0$ 
the third order polynomial (reduced master) equation  (\ref{4.masteqx})
has one and only one real solution $x = x_{*}$ which satisfies the inequality $x_{*} > 1$. 
}

{\bf Proof.} Let us fix $p_1 > 0$,  $p_2 > 0$ and consider the function $P(x)$
from (\ref{4.masteqx}). We get 
\begin{eqnarray}
P(0) = (1/2) p_1 p_2 > 0,  \label{4.x0} \\
P(1) = - (1/2) (p_1 + 1) (p_2 + 1) < 0,  \label{4.x11} \\
P(x) \to + \infty, \quad {\rm  as} \quad x  \to + \infty,  \label{4.xplus} \\
P(x) \to - \infty, \quad {\rm  as} \quad x   \to - \infty.  \label{4.xminus}
\end{eqnarray}
By using relations (\ref{4.x0}), (\ref{4.x11}), (\ref{4.xplus}), (\ref{4.xminus})
and Intermediate Value Theorem for our continuos function $P(x)$ we find 
that there exists at least one root $x_{*}$  of the polynomial $P(x)$
which obeys $x_{*} > 1$, and also there exist (at least) two other roots  
$x_{***}$, $x_{**}$ obeying $x_{***} < 0$ and $0 < x_{**} < 1$. 
The triple of roots $x_{***} < x_{**} < x_{*}$ is unique since otherwise 
we get that qubic polynomial has more than $3$ roots.
The  Proposition 2 is proved. 

Now we find explicit relations for  these three roots. By using
Mathematica we find three complex roots of the master equation
(\ref{4.masteqx}):
\begin{eqnarray}
  x_1  =  e^{2 i \pi/3} Y^{1/3} + e^{-2 i \pi/3} S Y^{-1/3} + 1/2,
 \label{4.x1} \\
 x_2  = e^{-2 i \pi/3} Y^{1/3} + e^{2 i \pi/3} S Y^{-1/3} + 1/2,
  \label{4.x2} \\
 x_3 =  Y^{1/3} +  S Y^{-1/3} + 1/2. 
\label{4.x3}
\end{eqnarray}

Here
\begin{equation}
Y = \sqrt{- {\cal R}}/(8 \times  3^{3/2}) + Z,
\label{4.Y}
\end{equation}
where
\begin{eqnarray} 
  {\cal R} =  64 p_1^3 p_2^3+ 96 p_1^3 p_2^2 + 96 p_1^2 p_2^3  
   \nonumber  \\
   + 48 p_1^3 p_2 + 240 p_1^2 p_2^2 + 48 p_1 p_2^3
   \nonumber  \\ 
   +8 p_1^3 + 168 p_1^2 p_2 + 168 p_1 p_2^2 + 8 p_2^3
       \nonumber  \\    
    +9 p_1^2 + 126 p_1 p_2 + 9 p_2^2,  
           \label{4.R}
 \end{eqnarray} 
and 
\begin{eqnarray} 
Z = (p_1+ p_2 +1)/8 > 1/8   \label{4.Z},\\
S = (4 p_1 p_2 +2 p_1 + 2 p_2 +3)/12 > 1/4.  \label{4.S}
 \end{eqnarray} 

Now we show that all these roots are real. 
By choosing the following value for the square root:
$\sqrt{-{\cal R}} = i \sqrt{{\cal R}} $ we obtain
 \begin{equation}
Y = i \frac{\sqrt{\cal R }}{8 \times 3^{3/2}} + Z  =  |Y| \exp{(i \alpha)},  
 \label{4.Y1}
 \end{equation}
where $0 < \alpha < \pi/2$.

Let us consider the third root
\begin{eqnarray} 
x_3  = Y^{1/3} +  S Y^{-1/3} + \frac{1}{2}  
\nonumber  \\
= |Y|^{1/3} e^{i \alpha/3} 
 +  S|Y|^{-1/3} e^{ - i \alpha/3} + \frac{1}{2}.
\label{4.x3R}
\end{eqnarray}

Here
\begin{eqnarray}
|Y| =  \sqrt{\frac{{\cal R}}{64 \times 27} + Z^2},  \label{4.modY} \\ 
\alpha = \arctan{\left(\frac{\sqrt{{\cal R}}}{8 Z \times 3^{3/2}}\right)},
\label{4.alpha}   
\end{eqnarray}
where $0< \alpha< \pi/2$.

Now we prove the following identity
\begin{equation}
|Y|^{1/3} = S|Y|^{-1/3},     \label{4.modYY}               
\end{equation}
which implies
\begin{equation}
x_3  =  2 |Y|^{1/3} \cos{(\alpha/3)} + \frac{1}{2}.
\label{4.x3cos}  
\end{equation}
In order order to prove this one should prove
\begin{equation}
|Y|^2 = S^3,  \label{4.modYS}                  
\end{equation}
or, equivalently,   
\begin{equation}
\frac{{\cal R}}{64 \times 27} + Z^2 =  S^3. 
\label{4.RS}
\end{equation}
But this identity is correct one. It was  verified by Mathematica.

Since $0 < \alpha/3 < \pi/6$ we get  
\begin{equation}
\cos{(\alpha/3)} > \frac{\sqrt{3}}{2}. 
\label{4.cos}
\end{equation}
But due to relations (\ref{4.Z}), (\ref{4.modY}) 
we obtain
\begin{equation}
|Y| > Z > \frac{1}{8}. 
\label{4.YZ}
\end{equation}
Combining (\ref{4.cos}) and (\ref{4.YZ}), we find
 \begin{eqnarray}
 x_3  =  2 |Y|^{1/3} \cos{(\alpha/3)} + \frac{1}{2} >  \nonumber \\
 2 \times \frac{1}{2} \times \frac{\sqrt{3}}{2} + \frac{1}{2} = 
  \frac{\sqrt{3}}{2} + \frac{1}{2} > 1. 
 \label{4.x3xstar}  
 \end{eqnarray}
 Hence, $x_{*} = x_3$. Thus, we have obtained the analytical relation 
 for the radius of photon sphere $R_0 = 2 \mu x_3$, with $x_3$ given 
 by relation (\ref{4.x3cos}).   
  
  Analogous consideration for other two roots gives us
  \begin{eqnarray}
   x_1  =  2 |Y|^{1/3} \cos{(\alpha/3 + 2 \pi/3)} + \frac{1}{2} < 0, 
   \label{4.x1xstar}  \\
   x_2  =  2 |Y|^{1/3} \cos{(\alpha/3 - 2 \pi/3)} + \frac{1}{2} < \frac{1}{2}, 
      \label{4.x2xstar} 
   \end{eqnarray}
    and $x_1 < x_2$, which imply $x_{***} = x_1$ and $x_{**} = x_2$ (see the proof of Proposition 2). 
    These inequalities   are in agreement with the results of numerical clculations for 
   certain solutions to (reduced) master equation (\ref{4.masteqx}) presented at Table 1.

          \begin{table} 
          \begin{center}
              \begin{tabular}{|c|c|c|c|c|}
              \hline
              $p_1$ & $p_2$ & $x_1$ & $x_2$ & $x_3$ \\
              \hline
              1 & 1 & -1 & 0.219 & 2.28 \\
              \hline
              2 & 1 & -1.43 & 0.262 & 2.67 \\
              \hline
              2 & 2 & -2 & 0.314 & 3.19 \\
              \hline
              3 & 1 & -1.781 & 0.28 & 3 \\
              \hline
              3 & 2 & -2.46 & 0.337 & 3.62 \\
              \hline
              3 & 3 & -3 & 0.363 & 4.14 \\
              \hline
              4 & 1 & -2.08 & 0.292 & 3.29 \\
              \hline
              4 & 2 & -2.851 & 0.351 & 4 \\
              \hline
              4 & 3 & -3.47 & 0.377 & 4.59 \\
              \hline
              4 & 4 & -4 & 0.391 & 5.11 \\
              \hline
              5 & 1 & -2.35 & 0.299 & 3.55 \\
              \hline
              5 & 2 & -3.2 & 0.359 & 4.34 \\
              \hline
              5 & 3 & -3.887 & 0.386 & 5 \\
              \hline
              5 & 4 & -4.47 & 0.401 & 5.57 \\
              \hline
              5 & 5 & -5 & 0.411 & 6.089 \\
              \hline
              \end{tabular}   
          \end{center}
      \caption{Examples of roots of reduced master equation 
      (\ref{4.masteqx}) for certain parameters $p_1$ and $p_2$. }
        \label{tab: x123}
        \end{table}
  
    Now we prove instability of circular photonic solution. To do that 
  we rewrite the effective potential (\ref{4.eff_pot}) with $k=0$ 
  by using dimesionless variables from  (\ref{4.xp}).
  We obtain 
  \begin{equation}
  U = \frac{\hat{L}^2}{(2 \mu)^2} F(x), 
  \label{4.UF}
  \end{equation}
 where 
 \begin{equation}
  F(x) = ( 1 + p_1/x)^{-2}( 1 + p_2/x)^{-2}( 1 - 1/x) x^{-2}.
    \label{4.Fx}
   \end{equation}
 
 For the first derivative of potential we get
 \begin{equation}
   \frac{d U}{d R} = \frac{\hat{L}^2}{(2 \mu)^3} \frac{dF}{dx},
     \label{4.dUdR}
    \end{equation}
 where 
 \begin{equation}
   \frac{ d F}{d x} = - \frac{(2 x^3-3 x^2 - 2 p_1 p_2 x - p_1 x - p_2 x + p_1 p_2)}{(x+p_1)^3 (x+ p_2)^3}
     \label{4.dFdx} .
  \end{equation}
 Here the reduced master equation  (\ref{4.masteqx}) is equivalent to  $\frac{dF}{dx} = 0$.
 
 For the second derivative of potential in the point of extremum ($\frac{d U}{d R} = 0$)  we obtain 
 \begin{equation}
    \frac{d^2 U}{d R^2} = \frac{\hat{L}^2}{(2 \mu)^4} \frac{d^2 F}{dx^2},
      \label{4.d2UdR2}
     \end{equation}
  where the second derivative of $F$ in the point of extremum ($\frac{dF}{dx} = 0$) reads
  \begin{equation}
    \frac{d^2F}{dx^2} = - \frac{(6x^2- 6x -2 p_1 p_2 - p_1  - p_2) }{(x+p_1)^3 (x+ p_2)^3}.
      \label{4.d2Fdx2}
       \end{equation}
 
 {\bf Proposition 3.} {\it For all $\mu > 0$, $P_1 > 0$,  $P_2 > 0$  
 the circular photonic geodesics given by unique solution  to master equation  
 (\ref{4.masteqR}) (or, equivalently $\left(\frac{d U}{d R}\right)_{R = R_0} = 0$)
  with $ R_0 = 2 \mu x_3 > 2 \mu$  
 are unstable due to inequality 
 \begin{equation}
 \left(\frac{d^2 U}{d R^2}\right)_{R = R_0} < 0.
 \label{4.d2UdR2neg}
        \end{equation}
 }
 
 {\bf Proof.} Due to relations  (\ref{4.d2UdR2}) and   (\ref{4.d2Fdx2}) to prove the 
 (\ref{4.d2UdR2neg}) one need to prove that
  \begin{equation}
   6 x^2 -6 x 
   - 2 p_1 p_2 - p_1 - p_2 > 0,
       \label{4.q_in}
  \end{equation}
  for the third root of equation (\ref{4.masteqx})
  given by (\ref{4.x3cos}).
    Rewriting (\ref{4.q_in}) as
    \begin{equation}
      6 x (x -1) >  2p_1 p_2 + p_1 + p_2 
      \label{4.q_in_1}
     \end{equation}
  and plugging the relation $x  =  2 |Y|^{1/3} \cos{(\alpha/3)} + \frac{1}{2}$,     
 we get 
 \begin{eqnarray}
 6 \left[2 |Y|^{1/3} \cos{(\alpha/3)} + \frac{1}{2}\right] \times \nonumber  \\
  \times\left[2 |Y|^{1/3} \cos{(\alpha/3)} - \frac{1}{2} \right]
 \nonumber \\
  > 2p_1 p_2 + p_1 + p_2  \qquad \qquad
 \label{4.q_in_2}
 \end{eqnarray}
  and hence 
 \begin{equation}
    24 |Y|^{2/3} (\cos{(\alpha/3)})^2  > 2p_1 p_2 + p_1 + p_2 + \frac{3}{2}.
   \label{4.q_in_3}
   \end{equation}
 Since $0 < \alpha < \pi/2$, we get $(\cos{(\alpha/3)})^2 > (\cos{(\pi/6)})^2 = 3/4$
 and hence 
 \begin{equation}
   24 |Y|^{2/3} (\cos{(\alpha/3)})^2  > 18 |Y|^{2/3} .
   \label{4.q_in_4}
   \end{equation}
 Thus, in order to prove   (\ref{4.q_in_3}) one need to prove 
 inequality 
  \begin{equation}
    18 |Y|^{2/3} > 2p_1 p_2 + p_1 + p_2 + \frac{3}{2}, 
   \label{4.q_in_5}
   \end{equation}
 for all $p_1 > 0$ and  $p_2 > 0$, or equivalently
    \begin{equation}
        (18)^3 |Y|^{2} > \left(2p_1 p_2 + p_1 + p_2 + \frac{3}{2}\right)^3. 
       \label{4.q_in_6}
       \end{equation}
  By using relations for $|Y|$ from (\ref{4.modY}) and $Z$ from 
  (\ref{4.Z}) we rewrite inequality (\ref{4.q_in_6}) under the consideration as 
  \begin{eqnarray}
  W = 27 ({\cal R} + 27 (p_1 + p_2 + 1)^2) \nonumber \\
   - (4 p_1 p_2 + 2 p_1 + 2 p_2 + 3)^3 > 0. 
         \label{4.q_in_7}
         \end{eqnarray}
   The substitution of ${\cal R}$ from (\ref{4.R}) into (\ref{4.q_in_7}) 
   gives us 
  \begin{equation}
    W = 26 (4 p_1 p_2 + 2 p_1 + 2 p_2 + 3)^3 > 0  
           \label{4.q_in_8}
           \end{equation}
 which is valid for all $p_1 > 0$ and  $p_2 > 0$. Thus, the proposition is proved. 
 
By using relations (\ref{4.d2UdR2}) and  (\ref{4.d2Fdx2}) we obtain by product a formula
for the second derivative of effective potential in the point of maximum $R = R_0 = 2 \mu x_3$
 \begin{equation}
    \left(\frac{d^2 U}{d R^2}\right)_0  = 
    - \frac{\hat{L}^2}{(2 \mu)^4} \frac{(6x_3^2- 6x_3 -2 p_1 p_2 - p_1  - p_2) }{(x_3 + p_1)^3 (x_3 + p_2)^3}.
      \label{4.d2UdR2x3}
     \end{equation}
 This quantity define the Lyapunov exponent corresponding to the unstable circular null geodesics
 \cite{CMBWZ}
     \begin{equation}
         \lambda_{0} = \frac{\sqrt{- \left(\frac{d^2 U}{d R^2}\right)_0}}{\sqrt{2 U(R_0)} } .
       \label{4.Lyap}
     \end{equation}
  It describes an  exponential  deviations of geodesics from circular one (in terms of time variable $t$) 
  for small   $ \delta R =  R - R_0$: $ \delta R \sim a \ e^{ \lambda_{0} t } $ ($a$ is constant), 
  which can be readily obtained from relations  (\ref{4.ELhat}) and (\ref{4.EqForRtrue}).
    
    Another important quantity is the angular velocity related to this unstable circular orbit
    \begin{equation}
     \Omega_{0} = \sqrt{\hat{U}(R_0)},
      \label{4.Omega}
    \end{equation}
    where here and in what follows we use a reduced effective potential  $\hat{U} = U/\hat{L}^2$ 
    ($\hat{L} \neq 0$) for $k = 0$.  From  (\ref{4.eff_pot}) we obtain 
          \bear 
              \hat{U} = \hat{U}(R)  \nonumber \\
             = (H_1(R))^{-2} (H_2(R))^{-2} \left(1-\frac{2\mu}{R}\right)  \frac{1}{R^2}.
          \label{4.eff_pot_red}
          \ear
     Relation (\ref{4.Omega}) just defines the angular rotation (in terms of time variable $t$) 
     for a circular  photonic orbit
      \begin{equation}
          \phi - \phi_0 = \pm \Omega_{0}(t- t_0),
            \label{4.angrot}
          \end{equation}
       where $\phi_0$ and $t_0$ are arbitrary constants. Equation  
       (\ref{4.angrot}) can be readily obtained from relations   
       (\ref{4.ELhat}) and $U(R_0) = \hat{E}^2$ (see (\ref{4.EqForR})).
   
   \subsection{Examples}
                 
   {\bf Example 1.}
   Let us  put 
   \begin{equation} 
   \label{4.P}
    p_1 = p_2 = p > 0.
   \end{equation}
   Then the dimensionless (reduced) master equation (\ref{4.masteqx}) reads
   \begin{equation} 
   \label{4.ZZ}
    (x + p) \left(  x^2 - \left( p + \frac{3}{2}  \right)  x   + \frac{1}{2} p \right) =0.
       \end{equation}
   
   After excluding the root $x_{1}= - p < 0$ we are led to  
   quadratic equation 
   \begin{equation} \label{4.QNM7x}
      x^2 - \left( p + \frac{3}{2}  \right)  x   + \frac{1}{2} p = 0 
   \end{equation}
   which yields other two solutions   
   \begin{eqnarray}
   x_{\pm}&=&\frac{1}{2} p+ \frac{3}{4} \pm \frac{1}{2}\sqrt{d}, \label{4.QNM8x}  \\
   d& =&  p^2 + p  + \frac{9}{4} > 0. \label{4.QNM8d}
   \end{eqnarray}
      It could be readily verified that   
   \begin{equation} 
     \label{4.QNM7zR0}
     x_3 = x_{+} > 1 > x_{-} = x_2 > 0
   \end{equation}
    for $p > 0$.
    
    {\bf Example 2.} Let $p_1 = q$, $p_2 = q - 2$, where $q > 2$. We obtain $x_3 = q$. 
   
   {\bf Example 3.} Let us put $p_1 = k_1 p$, $p_2 =  k_2 p $, where $k_1 > 0$ and $k_2 > 0$ are fixed constants 
   and $p > 0 $ is a variable under consideration. We obtain two asymptotic relations
   for the root $x_3$:  
   \begin{eqnarray}
      x_{3} = \frac{3}{2} + \frac{p_1 + p_2}{3} + O(p^2), \ {\rm as} \ p \to +0, \label{4.x3smallp}  \\
      x_{3} \sim \sqrt{p_1 p_2}, \ {\rm as} \ p \to + \infty. \label{4.x3bigp}
      \end{eqnarray}

     \section{BH shadow} 
      
       By using a standard consideration \cite{PTs,IBBKMNZ} one may find that there exists   
       a special solution that describes a spiral-like geodesic curve with an infinite ``winding'' angle. 
       This curve serves as a boundary between two classes of solutions (crossing and non-crossing the photon 
       sphere) and corresponds to the critical (or shadow ) angle 
       $\vartheta_{sh}\in (0, \pi/2)$, which is defined as the angle between the tangent to the spiral curve 
       at the point of observer $(R_{obs}, \phi_{obs})$ and the radial line.
      
      If a light ray is emitted from $(R_{obs}, \phi_{obs})$ at an angle smaller than the critical angle:
      \begin{equation}\label{b.104}
      \vartheta < \vartheta_{sh},
      \end{equation}
      the ray will traverse the photon sphere and 
      enter the event horizon and fall into the black hole.
      Conversely, if the emission angle satisfies
      \begin{equation}\label{b.105}
      \vartheta > \vartheta_{sh},
      \end{equation}
      the ray will not reach the photon sphere and will escape to infinity 
      (after a finite number of revolutions).

        $R_0$ corresponds to the point of maximum of the effective potential and hence
      \begin{equation}\label{b.107}
      \hat{U}(R_0) > \hat{U}(R) > 0
      \end{equation}
       for  $R > R_0$.

      The black hole shadow angle $\vartheta_{sh}$ can be determined 
      using a well-established relation \cite{PTs} (see also \cite{IBBKMNZ}):
        \begin{equation}\label{b.113}
      \sin\vartheta_{sh} = \sqrt{\frac{\hat{U}(R_{obs})}{\hat{U}(R_0)}}.
      \end{equation}
      Here $\hat{U}(R)$ is reduced effective potential defined in 
        (\ref{4.eff_pot_red}).      
       Therefore, the shadow angle is given by 
      \begin{equation}\label{b.114}
      \vartheta_{sh} = \arcsin\sqrt{\frac{\hat{U}(R_{obs})}{\hat{U}(R_0)}}, 
      \qquad 0< \vartheta_{sh}< \frac{\pi}{2},
      \end{equation}
      for all $R_{obs} > R_0$.  
      Here, $R_{obs}$ denotes the radial coordinate (position) of a point-like source or 
      light receiver (observer), while $R_0$ represents the radius of the photon sphere.

      For sufficiently large values of the observer's radial coordinate ($R_{obs}$)
      satisfying $R_{obs} \gg P$ and $R_{obs} \gg \mu$, the shadow angle can be determined
      using the following asymptotic formula:
      \begin{equation}\label{b.115}
         \vartheta_{sh} = \frac{b_0}{R_{obs} } + O ( 1/R^2_{obs}), 
      \end{equation}
      as $R_{obs} \to + \infty$. 
       Here
       \begin{equation}\label{b.116}
         b_0  = \frac{1}{\sqrt{\hat{U}(R_0)}}.
         \end{equation} 
      is critical impact parameter \cite{PTs}. 
      It follows from  (\ref{4.Omega}) that $b_{0} = (\Omega_0)^{-1}$.
      (Here the relation $c = 1$ for the speed of light is adopted.) 
       For dyonic BH under consideration    we obtain
        \begin{equation}\label{b.117}
        b_{0} = R_0 H_1(R_0) H_2(R_0) \left(1-\frac{2\mu}{R_0}\right)^{-1/2}.
       \end{equation}
     It is obvious, that $ b_{0} > R_0$ for all $\mu > 0$, $P_1 > 0$ and $P_2 > 0$.
     
     Let us fix $\mu > 0$. We get 
      \begin{equation}\label{b.118}
       b_{0} \to  3 \sqrt{3} \mu 
      \end{equation}
     in the (Schwarzschild) limit when $P_1 \to +0$, $P_2 \to +0$. In the case when
      $P_1 \to + \infty$, $P_2 \to + \infty$ and $P_1/P_2 = k = {\rm const}$
      we obtain the asymptotical relation 
      \begin{equation}\label{b.119}
            b_0 \sim R_0 \sim  \sqrt{P_1 P_2} \sim \sqrt{|Q_1| |Q_2|} 
      \end{equation}
      (see  (\ref{i4.3}) and (\ref{4.x3bigp}) above) and
      $ b_{0} \to + \infty$.

%%%%%%%%%%%%%%%%%%%n%%%%%%%%%%%%%%%%%%%%%%%%%%%%%%%%%%%%%%%%%%%%%
\section{Conclusions}
%%%%%%%%%%%%%%%%%%%%%%%%%%%%%%%%%%%%%%%%%%%%%%%%%%%%%%%%%%%%%%%%

In this paper we have explored a non-extremal black hole dyon
solution in  $4d$ gravitational model
with one scalar field and  one Abelian vector field.
The $4D$ model contains  dilatonic coupling $\lambda$ obeying
 $\lambda^2 = \frac{1}{2}$.
 
The black hole solutions are governed by two modulus functions, $H_1(R)$ and
$H_2(R)$, which depend on the parameters $P_1 > 0$ and $P_2 > 0$, respectively.

We have outlined several physical parameters of the solution, 
including the gravitational mass ($M$), the scalar charge ($Q_{\varphi}$), 
the Hawking temperature, and the Bekenstein-Hawking entropy associated with the black hole area. 
 
In this article, we have examined the circular null geodesics with a constant radius $R_0$,
 which corresponds to the radius of the photon sphere, and derived the master equation 
 for $R_0$. This equation is defined by a third-order polynomial  that depends on the parameters 
 of the black hole solution: $P_1$, $P_2$ and $ \mu  > 0$.

 We have proved the Proposition on existence and uniquenes of the solution to master equation
 obeying $R_0 > R_g = 2 \mu$. 
  We have also proved that circular null geodesics are  unstable.  
 
 Here we have also explored the BH shadow by presenting the shadow angle and 
 critical impact parameter.  
 
 For future investigations it  may be of interest to generalize the results obtained in this article to dyonic BH in nonlinear  electodynamics, see \cite{Br17} and references therein.

\vspace{5pt}

{\bf FUNDING}

For V.D.I.  this research was funded by RUDN University, scientific project number FSSF-2023-0003. 

%\newpage

\vspace{5pt}

{\bf CONFLICT OF INTERESTS}

The authors declare that they have no the conflicts of interests.

\end{document}